\documentclass[aps,pra,reprint]{revtex4-2}
\usepackage{graphicx, amsmath, amssymb, hyperref, xcolor}
\usepackage[caption=false]{subfig}
\allowdisplaybreaks[1] 

\begin{document}


\newcommand{\braket}[2]{{\left\langle #1 \middle| #2 \right\rangle}}
\newcommand{\bra}[1]{{\left\langle #1 \right|}}
\newcommand{\ket}[1]{{\left| #1 \right\rangle}}
\newcommand{\ketbra}[2]{{\left| #1 \middle\rangle \middle \langle #2 \right|}}
\newcommand{\fref}[1]{Fig.~\ref{#1}}


\title{Search by Lackadaisical Quantum Walk with Symmetry Breaking}

\author{Jacob Rapoza}
	\email{jacobrapoza@creighton.edu}
\author{Thomas G.~Wong}
	\email{thomaswong@creighton.edu}
	\affiliation{Department of Physics, Creighton University, 2500 California Plaza, Omaha, NE 68178}

\begin{abstract}
	The lackadaisical quantum walk is a lazy version of a discrete-time, coined quantum walk, where each vertex has a weighted self-loop that permits the walker to stay put. They have been used to speed up spatial search on a variety of graphs, including periodic lattices, strongly regular graphs, Johnson graphs, and the hypercube. In these prior works, the weights of the self-loops preserved the symmetries of the graphs. In this paper, we show that the self-loops can break all the symmetries of vertex-transitive graphs while providing the same computational speedups. Only the weight of the self-loop at the marked vertex matters, and the remaining self-loop weights can be chosen randomly, as long as they are small compared to the degree of the graph.
\end{abstract}

\maketitle


\section{Introduction}

The discrete-time, coined quantum walk is a quantum analogue of a discrete-time random walk, where a walker jumps between adjacent vertices of a graph in superposition. It was first proposed by Meyer as a quantum version of a cellular automaton \cite{Meyer1996a}, and he showed that for the evolution to be nontrivial, an internal degree of freedom was needed \cite{Meyer1996b}. Meyer identified the internal degree of freedom as spin and showed that the one-dimensional quantum walk was a discretization of the Dirac equation of relativistic quantum mechanics. Later, the internal degree of freedom was dubbed a ``coin'' in the context of a quantum walk \cite{Aharonov2001}, so that the quantum walk evolves by alternating between a quantum coin flip and a shift to adjacent vertices. The discrete-time, coined quantum walk has been used to design a variety of quantum algorithms, including algorithms for searching \cite{SKW2003}, solving element distinctness \cite{Ambainis2004}, and solving boolean formulas \cite{Childs2009NAND}. Furthermore, it is universal for quantum computing \cite{Lovett2010}, so any quantum circuit can be converted into a discrete-time quantum walk.

The lackadaisical quantum walk is a lazy version of this. It was introduced in \cite{Wong10} for searching the complete graph, which is the walk-formulation of Grover's unstructured search problem \cite{Grover1996}. In this initial work, $\ell$ integer self-loops were added to each vertex, with larger values of $\ell$ corresponding to greater laziness since there were more loops through which the walker could stay put. Later, the $\ell$ unweighted self-loops at each vertex were replaced by a single self-loop of real-valued weight $\ell$ at each vertex, such that if $\ell$ is an integer, it is equivalent to the original definition of $\ell$ integer self-loops per vertex \cite{Wong27}.

This generalization to real-valued weights led to speedups for spatial search on a variety of graphs, including the discrete torus with one marked vertex \cite{Wong28} and multiple marked vertices \cite{Saha2018,Nahimovs2019,Giri2019,deCarvalho2020,Saha2021}, periodic square lattices of arbitrary dimension \cite{Giri2020,Wong34}, strongly regular graphs \cite{Wong34}, Johnson graphs \cite{Wong34}, the hypercube \cite{Wong34}, regular locally arc-transitive graphs \cite{Hoyer2020}, the triangular lattice \cite{Nahimovs2021}, and the honeycomb lattice \cite{Nahimovs2021}. All of these graphs are vertex transitive, meaning they have symmetries such that each vertex has the same structure. Then, adding a self-loop of weight $\ell$ to each vertex preserves this symmetry, i.e., the graphs remain vertex transitive. Hanoi networks has also been explored \cite{Giri2019}, and although they are not vertex transitive, they do have some symmetry such that certain vertices have the same structure. This symmetry remains when adding a self-loop of weight $\ell$ to every vertex. In all these prior works, the self-loops preserved all the symmetries of the graphs. 

Lackadaisical quantum walks with nonhomogeneous weights were introduced for searching complete bipartite graphs \cite{Wong32}, where the self-loops in one partite set had one weight, and the self-loops in the other partite set had another weight. Regular complete bipartite graphs are vertex transitive, and although the nonhomogeneous weights broke this symmetry, not all the symmetry was broken, as vertices within a partite set still evolved identically. Irregular complete bipartite graphs are not vertex transitive, but vertices within a partite set have the same structure, and the nonhomogeneous weights retained this symmetry. Thus, in the regular case, the self-loops supported some of the symmetries of the graphs, and in the irregular case, they supported all the symmetries of the graph.

In this paper, we show that the self-loops can break all the symmetries of vertex-transitive graphs and still provide the same computational speedups. We show that only the weight at the marked vertex matters---all the other self-loops can be weighted randomly. In the next section, we define the quantum search algorithm by focusing on the complete graph, and we show that the speedup provided by the lackadaisical quantum walk remains when breaking the symmetry of the graph. Then, in Section III, we present similar findings for other vertex-transitive graphs, with search on periodic lattices suggesting that the random weights should be small compared to the degree of the graph. Finally, we conclude in Section IV.


\section{Complete Graph}

\begin{figure}
\begin{center}
	\includegraphics{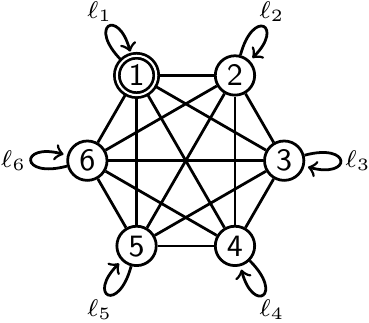}
	\caption{\label{fig:complete}A complete graph with $N = 6$ vertices and self-loops $\ell_1, \dots, \ell_6$. A vertex is marked, indicated by a double circle.}
\end{center}
\end{figure}

In this section, we revisit searching the complete graph of $N$ vertices using a lackadaisical quantum walk, except each self-loop can have a different weight. An example is shown in \fref{fig:complete}, where we have $N = 6$ vertices with self-loops of weight $\ell_1, \ell_2, \dots \ell_6$. This allows us to break the symmetries of the graph.

The system evolves in the Hilbert space $\mathcal{C}^N \otimes \mathcal{C}^N$ with basis vectors $\ket{u} \otimes \ket{v} = \ket{uv}$ denoting a walker at vertex $u$ pointing toward vertex $v$. The system begins in a uniform superposition over the vertices, and the amplitude at each vertex is distributed along the edges by weight:
\[ \ket{\psi(0)} = \frac{1}{\sqrt{N}} \sum_{i=1}^N \ket{i} \otimes \frac{1}{\sqrt{N + \ell_i - 1}} \left( \sum_{j \sim i} \ket{j} + \sqrt{\ell_i} \ket{i} \right). \]
We have access to an oracle $Q$ that we can query, and it negates the amplitudes at the ``marked'' vertex. Let $\ket{a} \in \{ \ket{1}, \dots, \ket{N} \}$ denote the marked vertex. Then,
\[ Q = (I_N - 2\ketbra{a}{a}) \otimes I_N. \]
The quantum walk consists of a coin flip $C$ and a shift $S$. We use the ``Grover diffusion coin'' with a weighted self-loop \cite{Wong27}, defined as
\[ C = \sum_{i=1}^N \big[ \ketbra{i}{i} \otimes (2 \ketbra{s_i}{s_i} - I_N) \big], \]
where
\[ \ket{s_i} = \frac{1}{\sqrt{N + \ell_i - 1}} \left( \sum_{j \sim i} \ket{j} + \sqrt{\ell_i} \ket{i} \right). \]
The shift causes a particle to jump and turn around, i.e.,
\[ S \ket{uv} = \ket{vu}. \]
The search algorithm evolves by repeated applications of
\[ U = SCQ, \]
which queries the oracle $Q$ and then takes a step of the quantum walk $SC$. So, $\ket{\psi(t)} = U^t \ket{\psi(0)}$.

\begin{figure}
\begin{center}
       	\subfloat[] {
		\includegraphics{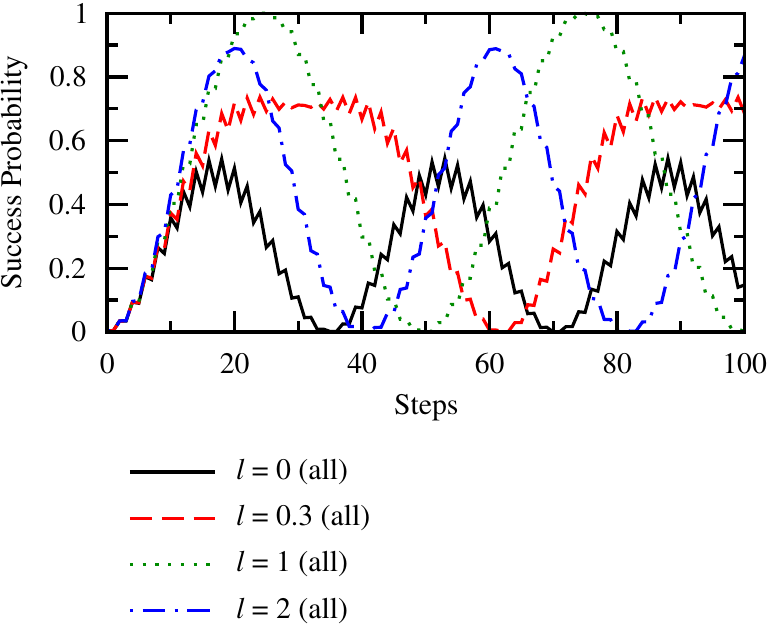}
		\label{fig:complete_256}
	}
    
	\subfloat[] {
		\includegraphics{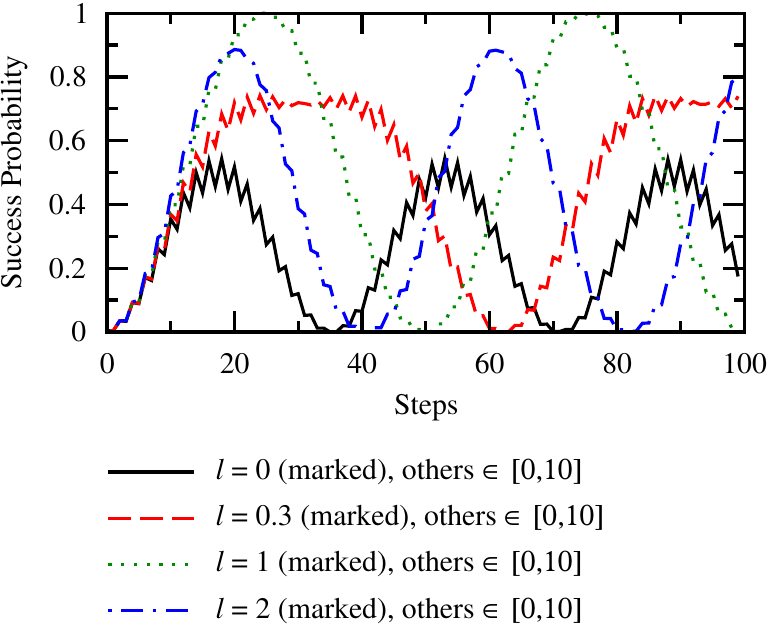}
		\label{fig:complete_256_rand}
	}
	\caption{Search on the complete graph of $N = 256$ vertices, which has degree $255$. In (a), every self-loop has weight $\ell$. In (b), the self-loop at the marked vertex has weight $\ell$, while the rest are chosen uniformly at random in the interval $[0,10]$.}
\end{center}
\end{figure}

In previous research \cite{Wong27}, all the self-loops had the same weight, i.e., $\ell_1 = \dots = \ell_N = \ell$, and it was shown that the search algorithm behaved differently for different values of $\ell$. This is shown in \fref{fig:complete_256} for search on the complete graph with $N = 256$ vertices. The black solid curve corresponds to $\ell = 0$, which is the loopless algorithm. The success probability (i.e., the probability at the marked vertex) starts at 1/256, and as $U$ is repeatedly applied, it rises to a success probability of $1/2$ after $\pi\sqrt{N}/2\sqrt{2} \approx 18$ steps. Then, the success probability decreases again in a quasi-periodic manner. The dashed red curve is $\ell = 0.3$, and the success probability now reaches a value of 0.71. With $\ell = 1$, corresponding to the dotted green curve, the success probability now reaches 1. Finally, when $\ell = 2$, shown in the dot-dashed blue curve, the success probability reaches 0.89. Thus, the optimal value of $\ell$ that maximally boosts the success probability is $\ell = 1$. 

Analytically, it was shown in Section 6 of \cite{Wong27} that, for large $N$, the success probability at time $t$ for the homogeneous lackadaisical quantum walk is
\begin{align*}
	p_h(t)
		&= \left[ \frac{[1-\cos(\alpha t)] \sqrt{\ell(N-1)}}{(\ell+1)\sqrt{N+\ell-2}} \right]^2 \\
	       	&\quad + \left[ \frac{\sqrt{(2N+\ell-3)(\ell+1)} \sin(\alpha t)}{2(\ell+1)\sqrt{N+\ell-2}} \right]^2,
\end{align*}
where
\[ \alpha = \sin^{-1} \left( \frac{\sqrt{(2N+\ell-3)(\ell+1)}}{N+\ell-1} \right). \]
Siimplifying these further for large $N$:
\begin{align}
	p_h(t)
		&= \frac{\ell \left[1 - \cos(\alpha t) \right]^2}{(\ell+1)^2} + \frac{\sin^2(\alpha t)}{2(\ell+1)} \nonumber \\
		&= \frac{8l \sin^4(\alpha t/2) + (\ell+1) \sin^2(\alpha t)}{2(\ell+1)^2}, \label{eq:prob}
\end{align}
and
\begin{equation}
	\label{eq:alpha}
	\alpha = \sin^{-1}  \left( \sqrt{\frac{2(\ell+1)}{N}} \right).
\end{equation}
Also from \cite{Wong27}, the success probability $p_h(t)$ reaches a peak of
\[ p_* = \begin{cases}
	\frac{1}{2(1-\ell)}, & \ell < 1/3, \\
	\frac{4\ell}{(\ell+1)^2}, & \ell \ge 1/3,\ \ell = o(N), \\
	\frac{16+9c}{4c(c+1)} \frac{1}{N}, & \ell = cN, \\
	\frac{9}{4\ell}, & \ell = \omega(N), \\
\end{cases} \]
at time
\[ t_* = \begin{cases}
	\frac{\cos^{-1} \left( \frac{2\ell}{\ell-1} \right)}{\sqrt{2(\ell+1)}} \sqrt{N}, & \ell < 1/3, \\
	\frac{\pi}{\sqrt{2(\ell+1)}} \sqrt{N}, & l \ge 1/3,\ \ell = o(N), \\
	\frac{\pi}{\sin^{-1} \left( \frac{\sqrt{c(c+2)}}{c+1} \right)}, & \ell = cN, \\
	2, & \ell = \omega(N). \\
\end{cases} \]
For example, when $\ell = 1$, the success probability reaches $p_* = 1$ at time $t_* = \pi\sqrt{N}/2$. Or, when $\ell = 0$, the success probability reaches $p_* = 1/2$ at time $t_* = \pi\sqrt{N}/2\sqrt{2}$.

Now, say the self-loop at the marked vertex has weight $\ell$ while the remaining $255$ self-loops have weights chosen uniformly at random between 0 and 10. This breaks all the symmetries of the graph, as each vertex has a different structure. In \fref{fig:complete_256_rand}, we plot the success probability with the same values of $\ell$ as in \fref{fig:complete_256}, i.e., $\ell = 0, 0.3, 1, 2$. Comparing these figures, the success probabilities evolve nearly identically, indicating that only the weight of the self-loop at the marked vertex matters.

Let us prove that only the weight at the marked vertex affects the search algorithm, asymptotically, for the case where some self-loops have one weight and the remaining vertices have another weight. This breaks the vertex transitivity of the graph since the vertices no longer all have the same structure. Hence, it proves that some of the symmetry of the graph can be broken while preserving the speedup, as for the regular complete bipartite graph \cite{Wong32}. Proving the general case with every self-loop taking a different value, as in \fref{fig:complete_256_rand}, is open.

\begin{figure}
\begin{center}
	\subfloat[] {
		\label{fig:complete_M}
		\includegraphics{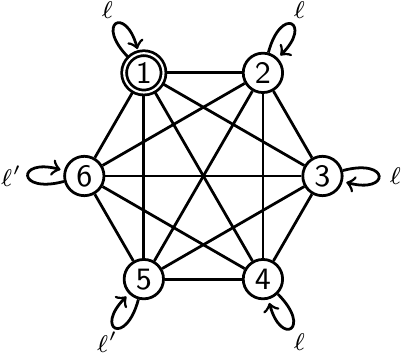}
	}
	\subfloat[] {
		\label{fig:complete_M_colored}
		\includegraphics{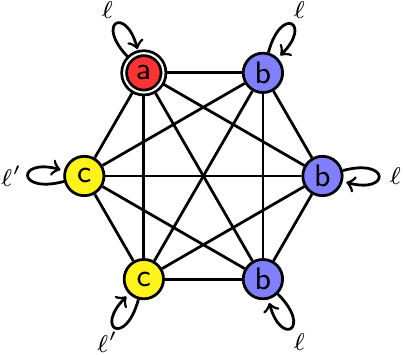}
	}
	\caption{(a) A complete graph with $N = 6$ vertices, where $M = 4$ vertices have self-loops of weight $\ell$ and $(N-M) = 2$ vertices have self-loops of weight $\ell'$. A vertex is marked, indicated by a double circle. (b) The same graph, but with identically evolving vertices identically labeled and colored.}
\end{center}
\end{figure}

To begin the proof, we assume $M$ of the vertices have self-loops of weight $\ell$, and the remaining $(N-M)$ vertices have self-loops of weight $\ell'$. Without loss of generality, we take the $M$ vertices with weights $\ell$ to have labels $1, 2, \dots, M$, and we take the remaining $(N-M)$ vertices with weights $\ell'$ to have labels $M+1, M+2, \dots, N$. Then, the initial state of the system is
\begin{align}
	\ket{\psi(0)} 
		&= \frac{1}{\sqrt{N}} \Bigg[ \sum_{i = 1}^{M} \ket{i} \otimes \frac{1}{\sqrt{N+\ell-1}} \left( \sum_{j \ne i} \ket{j} + \sqrt{\ell} \ket{i} \right) \nonumber \\
		&\quad+ \sum_{i = M+1}^{N} \ket{i} \otimes \frac{1}{\sqrt{N+\ell'-1}} \left( \sum_{j \ne i} \ket{j} + \sqrt{\ell'} \ket{i} \right) \Bigg]. \label{eq:s}
\end{align}
Without loss of generality, we assume that the marked vertex is among the $M$ vertices with self-loop weight $\ell$. An example with $N = 6$ vertices and $M = 4$ is shown in \fref{fig:complete_M}.

With these assumptions, many of the vertices evolve identically to each other. In \fref{fig:complete_M_colored}, we have labeled and colored identically-evolving vertices the same. There are only three types of vertices: The marked vertex is labeled $a$ and is red, the other $(M-1)$ vertices with self-loops $\ell$ are labeled $b$ and are blue, and the $(N-M)$ vertices with self-loops $\ell'$ are labeled $c$ and are yellow. Taking into account the direction that a walker at each vertex can point, the system evolves in a 9D subspace spanned by
\begin{align*}
	&\ket{aa} = \ket{a} \otimes \ket{a}, \\
	&\ket{ab} = \ket{a} \otimes \frac{1}{\sqrt{M-1}} \sum_b \ket{b}, \\
	&\ket{ac} = \ket{a} \otimes \frac{1}{\sqrt{N-M}} \sum_c \ket{c}, \\
	&\ket{ba} = \frac{1}{\sqrt{M-1}} \sum_b \ket{b} \otimes \ket{a}, \\
	&\ket{bb} = \frac{1}{\sqrt{M-1}} \sum_b \ket{b} \otimes \frac{1}{\sqrt{M+\ell-2}} \left(\sum_{b' \ne b} \ket{b'} + \sqrt{\ell} \ket{b} \right), \\
	&\ket{bc} = \frac{1}{\sqrt{M-1}} \sum_b \ket{b} \otimes \frac{1}{\sqrt{N-M}} \sum_{c} \ket{c}, \\
	&\ket{ca} = \frac{1}{\sqrt{N-M}} \sum_{c} \ket{c} \otimes \ket{a}, \\
	&\ket{cb} = \frac{1}{\sqrt{N-M}} \sum_{c} \ket{c} \otimes \frac{1}{\sqrt{M-1}} \sum_b \ket{b}, \\
	&\ket{cc} = \frac{1}{\sqrt{N-M}} \sum_{c} \ket{c} \\
	&\quad\quad\quad \otimes \frac{1}{\sqrt{N+\ell'-M-1}} \left( \sum_{c' \ne c} \ket{c'} + \sqrt{\ell'} \ket{c} \right).
\end{align*}
Then in this $\{ \ket{aa}, \ket{ab}, \dots, \ket{cc} \}$ basis, the initial state \eqref{eq:s} is
\begin{align}
	\ket{\psi(0)} 
	    	&= \frac{1}{\sqrt{N}} \Bigg( \sqrt{\frac{\ell}{N+\ell-1}} \ket{aa} + \sqrt\frac{{M-1}}{{N+\ell-1}} \ket{ab} \nonumber \\ 
		&\quad+ \sqrt\frac{N-M}{N+\ell-1} \ket{ac} + \sqrt\frac{M-1}{N+\ell-1} \ket{ba} \nonumber \\ 
		&\quad+ \sqrt\frac{{(M-1) (M+\ell-2)}}{{N+\ell-1}} \ket{bb} \nonumber \\
		&\quad+ \sqrt\frac{{(M-1) (N-M)}}{{N+\ell-1}} \ket{bc} \nonumber \\
		&\quad+ \sqrt\frac{{N-M}}{{N+\ell'-1}} \ket{ca} +\sqrt\frac{{(M-1) (N-M)}}{{N+\ell'-1}} \ket{cb} \nonumber \\ 
		&\quad+ \sqrt\frac{{(N-M) (N-M+\ell'-1)}}{{N+\ell'-1}} \ket{cc} \Bigg), \label{eq:s_subspace}
\end{align}
and the search operator $U = SCQ$ is
\begin{widetext}
\begin{equation}
	\label{eq:U}
	U = \begin{pmatrix}
		\frac{N-\ell-1}{N+\ell-1} & -\frac{2 \sqrt{\ell M_1}}{N+\ell-1} & -\frac{2 \sqrt{\ell} \sqrt{N_M}}{N+\ell-1} & 0 & 0 & 0 & 0 & 0 & 0 \\
		0 & 0 & 0 & \frac{-N-\ell+3}{N+\ell-1} & \frac{2 \sqrt{M_{\ell}}}{N+\ell-1} & \frac{2 \sqrt{N_M}}{N+\ell-1} & 0 & 0 & 0 \\
		0 & 0 & 0 & 0 & 0 & 0 & \frac{-N-\ell'+3}{N+\ell'-1} & \frac{2 \sqrt{M_1}}{N+\ell'-1} & \frac{2 \sqrt{N_{M\ell'}}}{N+\ell'-1} \\
		-\frac{2 \sqrt{\ell M_1}}{N+\ell-1} & \frac{N_{2M\ell}}{N+\ell-1} & -\frac{2 \sqrt{N_M M_1}}{N+\ell-1} & 0 & 0 & 0 & 0 & 0 & 0 \\
		0 & 0 & 0 & \frac{2 \sqrt{M_{\ell}}}{N+\ell-1} & -\frac{N_{-2M\ell} + 2}{N+\ell-1} & \frac{2 \sqrt{M_{\ell} N_M}}{N+\ell-1} & 0 & 0 & 0 \\
		0 & 0 & 0 & 0 & 0 & 0 & \frac{2 \sqrt{M_1}}{N+\ell'-1} & -\frac{N_{2M\ell'}}{N+\ell'-1} & \frac{2 \sqrt{M_1 N_{M\ell'}}}{N+\ell'-1} \\
		-\frac{2 \sqrt{\ell N_M}}{N+\ell-1} & -\frac{2 \sqrt{N_M M_1}}{N+\ell-1} & -\frac{N_{-2M\ell}}{N+\ell-1} & 0 & 0 & 0 & 0 & 0 & 0 \\
		0 & 0 & 0 & \frac{2 \sqrt{N_M}}{N+\ell-1} & \frac{2 \sqrt{M_{\ell} N_M}}{N+\ell-1} & \frac{N_{-2M\ell}}{N+\ell-1} & 0 & 0 & 0 \\
		0 & 0 & 0 & 0 & 0 & 0 & \frac{2 \sqrt{N_{M\ell'}}}{N+\ell'-1} & \frac{2 \sqrt{M_1 N_{M\ell'}}}{N+\ell'-1} & \frac{N_{-2M\ell'}}{N+\ell'-1} 
	\end{pmatrix},
\end{equation}
\end{widetext}
where
\begin{align*}
	&M_1 = M-1, \\
	&M_{\ell} = M+\ell-2, \\
	&N_M = N-M, \\
	&N_{M\ell'} = N-M+\ell'-1, \\
	&N_{2M\ell} = N-2M+\ell+1, \\
	&N_{2M\ell'} = N-2M+\ell'+1, \\
	&N_{-2M\ell} = N-2M-\ell+1, \\
	&N_{-2M\ell'} = N-2M+\ell'-1.
\end{align*}

To find the evolution of the system, we want to find the eigenvectors and eigenvalues of $U$ \eqref{eq:U}. Then, we can express the initial state \eqref{eq:s_subspace} as a linear combination of these eigenvectors, and the state of the system at time $t$ is obtained by simply multiplying each eigenvector by its eigenvalue $t$ times. This is difficult to do exactly, but it can be done asymptotically. In the next subsection, we do this assuming $N$ is the dominant variable. In the subsection after that, we assume $M$ scales with $N$, so for large $N$, there is also an asymptotic contribution from $M$. In both of these cases, we will prove that the success probability evolves the same as the homogeneous lackadaisical quantum walk in \eqref{eq:prob}. Before continuing on to other graphs, we end this section on the complete graph with a third subsection, showing that another reasonable initial state yields the same evolution, asymptotically.


\subsection{Large $N$}

In this subsection, we assume $N$ is the dominant variable, e.g., $N - M + \ell' - 1 \approx N$. That is, in little-o notation, $M = o(N)$. Then, for large $N$, the initial state \eqref{eq:s_subspace} becomes
\begin{equation}
	\label{eq:s_largeN}
	\ket{\psi(0)} = \ket{cc}.
\end{equation}
Then, as shown in Appendix~\ref{appendix:largeN} using degenerate perturbation theory, the (unnormalized) eigenvectors and eigenvalues of $U$ \eqref{eq:U} for large $N$ are
\begin{widetext}
\begin{align}
	& \ket{\Psi_1}=[0,0,0,0,1,0,0,0,0]^\intercal, \quad \lambda_1 = -1, \nonumber \\
	& \ket{\Psi_2}=\frac{1}{\sqrt{2}}[0,0,1,0,0,0,1,0,0]^\intercal, \quad \lambda_2 = -1, \nonumber \\
	& \ket{\Psi_3}=\frac{1}{\sqrt{2}}[0,i,0,1,0,-i,0,-1,0]^\intercal, \quad \lambda_3 = i - \frac{1}{\sqrt{N}}, \nonumber \\
	& \ket{\Psi_4}=\frac{1}{\sqrt{2}}[0,i,0,1,0,i,0,1,0]^\intercal, \quad \lambda_4 = i + \frac{1}{\sqrt{N}}, \nonumber \\
	& \ket{\Psi_5}=\frac{1}{\sqrt{2}}[0,-i,0,1,0,i,0,-1,0]^\intercal, \quad \lambda_5 = -i - \frac{1}{\sqrt{N}}, \label{eq:eigenvectors_largeN} \\
	& \ket{\Psi_6}=\frac{1}{\sqrt{2}}[0,-i,0,1,0,-i,0,1,0]^\intercal, \quad \lambda_6 = -i + \frac{1}{\sqrt{N}}, \nonumber \\
	& \ket{\Psi_7}=\left[ 1,0,i\sqrt{\frac{\ell+1}{2\ell}},0,0,0,-i\sqrt{\frac{\ell+1}{2\ell}},0,\frac{1}{\sqrt{\ell}} \right]^\intercal, \quad \lambda_7 = e^{-i\alpha}, \nonumber \\
	& \ket{\Psi_8}=\left[1,0,-i\sqrt{\frac{\ell+1}{2\ell}},0,0,0,i\sqrt{\frac{\ell+1}{2\ell}},0,\frac{1}{\sqrt{\ell}}\right]^\intercal, \quad \lambda_8 = e^{i\alpha}, \nonumber \\
	& \ket{\Psi_9}=[1,0,0,0,0,0,0,0,-\sqrt{\ell}]^\intercal, \quad \lambda_9 = 1 \nonumber,
\end{align}
\end{widetext}
where $\alpha$ is defined in \eqref{eq:alpha}.

To find the evolution of the system for large $N$, we express $\ket{\psi(0)}$ \eqref{eq:s_largeN} as a linear combination of the approximate eigenvectors of $U$ that we just found \eqref{eq:eigenvectors_largeN}:
\begin{align*}
	\ket{\psi(0)} 
		&= a\ket{\Psi_1}+b\ket{\Psi_2}+c\ket{\Psi_3}+d\ket{\Psi_4}+e\ket{\Psi_5}\\
		&\quad+ f\ket{\Psi_6}+g\ket{\Psi_7}+h\ket{\Psi_8}+i\ket{\Psi_9},
\end{align*}
where
\begin{align*}
	& a=b=c=d=e=f=0, \\
	& g=h=\frac{\sqrt{\ell}}{2(\ell+1)}, \\
	& i=-\frac{\sqrt{\ell}}{\ell+1}.
\end{align*}
That is,
\[ \ket{\psi(0)} = \frac{\sqrt{\ell}}{2(\ell+1)} \ket{\Psi_7} + \frac{\sqrt{\ell}}{2(\ell+1)} \ket{\Psi_8} - \frac{\sqrt{\ell}}{\ell+1} \ket{\Psi_9}. \]
Applying $U$ to this multiplies each eigenvector by its eigenvalue, so the state at time $t$ is
\begin{align*}
	\ket{\psi(t)}
		&= U^t \ket{\psi(0)} \\
		&= \frac{\sqrt{\ell}}{2(\ell+1)}e^{-i \alpha t}\ket{\Psi_7}+\frac{\sqrt{\ell}}{2(\ell+1)}e^{i \alpha t}\ket{\Psi_8} \\
		&\quad -\frac{\sqrt{\ell}}{\ell+1}\ket{\Psi_9}.
\end{align*}
Substituting in for the $\ket{\Psi_i}$'s, the state in the $\{ \ket{aa},\ket{ab}, \dots, \ket{cc} \}$ basis is
\begin{align}
	\ket{\psi(t)}
		&= \Bigg[ \frac{\sqrt{\ell}}{\ell+1} [\cos(\alpha t) - 1], 0, \frac{\sin(\alpha t)}{\sqrt{2(\ell+1)}}, 0, 0, 0, \label{eq:psi_largeN} \\
		&\quad\quad -\frac{\sin(\alpha t)}{\sqrt{2(\ell+1)}} , 0, \frac{\ell + \cos(\alpha t)}{\ell+1} ] \Bigg]^\intercal. \nonumber
\end{align}
The success probability with respect to time is the sum of the squares of the amplitudes of $\ket{aa}$, $\ket{ab}$, and $\ket{ac}$, which is
\begin{align}
	p(t) 
		&= \frac{\ell}{(\ell+1)^2} \left[ \cos(\alpha t) - 1 \right]^2 + \frac{\sin^2(\alpha t)}{2(\ell+1)} \nonumber \\
		&= \frac{8\ell\sin ^4(\frac{\alpha t}{2})+(\ell+1) \sin^2(\alpha t)}{2(\ell+1)^2}. \label{eq:prob_largeN}
\end{align}
This is exactly the same success probability as the homogeneous case \eqref{eq:prob}. So, asymptotically, the nonhomogeneous lackadaisical quantum walk evolves the same as the homogeneous lackadaisical quantum walk with weight $\ell$.


\subsection{Large $N$ and $M$}

In this subsection, we assume $M$ scales with $N$. That is, in big-Theta notation, $M = \Theta(N)$. Then, for large $N$, $M$ becomes a constant multiple of $N$. For example, if one-fourth of the self-loops have weight $\ell$ and three-fourths of the self-loops have weight $\ell'$, then $M = (1/4)N$. So in general, for large $N$, we can write $M = kN$ for some constant $k$. Then, for example, $N - M + \ell' - 1 = N - kN = (1-k)N$ for large $N$.

Then, for large $N$, the initial state \eqref{eq:s_subspace} becomes
\begin{align}
    \ket{\psi(0)} 
        &= k \ket{bb} + \sqrt{k(1-k)} \ket{bc} \nonumber \\
        &\quad + \sqrt{k(1-k)} \ket{cb} + (1-k) \ket{cc}. \label{eq:s_largeM}
\end{align}
As shown in Appendix~\ref{appendix:largeM}, using degenerate perturbation theory, the (unnormalized) eigenvectors and eigenvalues of $U$ \eqref{eq:U} are asymptotically
\begin{widetext}
\begin{align}
	\ket{\Psi_1}
		&=\frac{1}{\sqrt{2}} \left[ 0,\sqrt{k},\sqrt{1-k},\sqrt{k},0,0,\sqrt{1-k},0,0 \right]^\intercal, \quad \lambda_1 = -1, \nonumber \\
	\ket{\Psi_2}
		&=\left[ 0,0,0,0,1-k,-\sqrt{k(1-k)},0,-\sqrt{k(1-k)},k \right]^\intercal, \quad \lambda_2 = -1, \nonumber \\
	\ket{\Psi_3}
		&= \frac{1}{\sqrt{2}} \bigg[ 0,-i\sqrt{1-k},i\sqrt{k},-\sqrt{1-k},-(1+i)\sqrt{k(1-k)},(1+i)\left(k-\frac{1}{2}-\frac{i}{2}\right),\sqrt{k},(1+i)\left(k-\frac{1}{2}+\frac{i}{2}\right), \nonumber \\
		&\quad\quad\quad\quad (1+i)\sqrt{k(1-k)} \bigg]^\intercal, \quad \lambda_3 = ie^{-i\phi}, \nonumber \\
	\ket{\Psi_4}
		&= \frac{1}{\sqrt{2}} \bigg[ 0,-i\sqrt{1-k},i\sqrt{k},-\sqrt{1-k},(1+i)\sqrt{k(1-k)},-(1+i)\left(k-\frac{1}{2}-\frac{i}{2}\right),\sqrt{k},-(1+i)\left(k-\frac{1}{2}+\frac{i}{2}\right), \nonumber\\
		&\quad\quad\quad\quad -(1+i)\sqrt{k(1-k)} \bigg]^\intercal, \quad \lambda_4 = ie^{i\phi}, \nonumber \\
	\ket{\Psi_5}
		&= \frac{1}{\sqrt{2}} \bigg[ 0,i\sqrt{1-k},-i\sqrt{k},-\sqrt{1-k},-(1-i)\sqrt{k(1-k)},(1-i)\left(k-\frac{1}{2}+\frac{i}{2}\right),\sqrt{k},(1-i)\left(k-\frac{1}{2}-\frac{i}{2}\right), \label{eq:eigenvectors_largeM} \\
		&\quad\quad\quad\quad (1-i)\sqrt{k(1-k)} \bigg]^\intercal, \quad \lambda_5 = -ie^{i\phi}, \nonumber \\
	\ket{\Psi_6}
		&= \frac{1}{\sqrt{2}} \bigg[ 0,i\sqrt{1-k},-i\sqrt{k},-\sqrt{1-k},(1-i)\sqrt{k(1-k)},-(1-i)\left(k-\frac{1}{2}+\frac{i}{2}\right),\sqrt{k},-(1-i)\left(k-\frac{1}{2}-\frac{i}{2}\right), \nonumber \\
		&\quad\quad\quad\quad -(1-i)\sqrt{k(1-k)} \bigg]^\intercal, \quad \lambda_6 = -ie^{-i\phi}, \nonumber \\
	\ket{\Psi_7}
		&=\left[ 1,i\sqrt{\frac{k(\ell+1)}{2\ell}}, i\sqrt{\frac{(1-k)(\ell+1)}{2\ell}},-i\sqrt{\frac{k(\ell+1)}{2\ell}},\frac{k}{\sqrt{\ell}},\sqrt{\frac{k(1-k)}{\ell}},-i\sqrt{\frac{(1-k)(\ell+1)}{2\ell}},\sqrt{\frac{k(1-k)}{\ell}},\frac{1-k}{\sqrt{\ell}} \right]^\intercal, \nonumber \\
		&\quad\quad\quad\quad \lambda_7 = e^{-i\alpha}, \nonumber \\
	\ket{\Psi_8}
		&=\left[ 1,-i\sqrt{\frac{k(\ell+1)}{2\ell}}, -i\sqrt{\frac{(1-k)(\ell+1)}{2\ell}},i\sqrt{\frac{k(\ell+1)}{2\ell}},\frac{k}{\sqrt{\ell}},\sqrt{\frac{k(1-k)}{\ell}},i\sqrt{\frac{(1-k)(\ell+1)}{2\ell}},\sqrt{\frac{k(1-k)}{\ell}},\frac{1-k}{\sqrt{\ell}} \right]^\intercal, \nonumber \\
		&\quad\quad\quad\quad \lambda_8 = e^{i\alpha}, \nonumber \\
	\ket{\Psi_9}
		&=\left[ 1,0,0,0,-k\sqrt{\ell},-\sqrt{k(1-k)\ell},0,-\sqrt{k(1-k)\ell},-(1-k)\sqrt{\ell} \right]^\intercal, \quad \lambda_9 = 1, \nonumber
\end{align}
\end{widetext}
where
\[ \phi = \sin^{-1} \left( \frac{1}{\sqrt{N}} \right), \]
and $\alpha$ is defined in \eqref{eq:alpha}.
As before, we express the initial state \eqref{eq:s_largeM} as a linear combination of the eigenvectors \eqref{eq:eigenvectors_largeM}:
\begin{align*}
	\ket{\psi(0)} 
		&= a\ket{\Psi_1}+b\ket{\Psi_2}+c\ket{\Psi_3}+d\ket{\Psi_4}+e\ket{\Psi_5}\\
		&\quad+ f\ket{\Psi_6}+g\ket{\Psi_7}+h\ket{\Psi_8}+i\ket{\Psi_9},
\end{align*}
where
\begin{align*}
	& a = b = c = d = e = f = 0, \\
	& g = h = \frac{\sqrt{\ell}}{2(\ell+1)}, \\
	& i = -\frac{\sqrt{\ell}}{\ell+1}.
\end{align*}
In other words, for large $N$,
\[ \ket{\psi(0)} = \frac{\sqrt{\ell}}{2(\ell+1)} \ket{\Psi_7} + \frac{\sqrt{\ell}}{2(\ell+1)} \ket{\Psi_8} - \frac{\sqrt{\ell}}{\ell+1} \ket{\Psi_9}. \]
Applying $U$ then multiplies each eigenvector by its eigenvalue, so the state $\ket{\psi(0)}$ after $t$ applications is 
\begin{align*}
	\ket{\psi(t)}
		&= U^t \ket{\psi(0)} \\
		&= \frac{\sqrt{\ell}}{2(\ell+1)} e^{-i\alpha t} \ket{\Psi_7} + \frac{\sqrt{\ell}}{2(\ell+1)} e^{i\alpha t} \ket{\Psi_8} \\
		&\quad - \frac{\sqrt{\ell}}{\ell+1} (1)^t \ket{\Psi_9} \\
		&= \Bigg[ \frac{\sqrt{\ell}}{\ell+1} [\cos(\alpha t) - 1], \sqrt{\frac{k}{2(\ell+1)}} \sin(\alpha t), \\
		&\quad \sqrt{\frac{1-k}{2(\ell+1)}} \sin(\alpha t), -\sqrt{\frac{k}{2(\ell+1)}} \sin(\alpha t), \\
		&\quad \frac{k}{\ell+1} [\ell+\cos(\alpha t)], \frac{\sqrt{k(1-k)}}{\ell+1} [\ell + \cos(\alpha t)], \\
		&\quad - \sqrt{\frac{1-k}{2(\ell+1)}} \sin(\alpha t), \frac{\sqrt{k(1-k)}}{\ell+1} [\ell + \cos(\alpha t)], \\
		&\quad \frac{1-k}{\ell+1} [\ell + \cos(\alpha t)] \Bigg]^\intercal.
\end{align*}
Note when $k \to 0$, the amplitudes involving $b$ vertices (i.e., $\ket{ab}$, $\ket{ba}$, $\ket{bb}$, $\ket{bc}$, and $\ket{cb}$) all go to zero, and we get \eqref{eq:psi_largeN} from the previous section where $M$ is small compared to $N$. This is because for small $M$ and large $N$, the overwhelming majority of vertices are $c$ vertices, and the $b$ vertices do not play a significant role. In contrast, for large $M$, a significant number of vertices are also $b$ vertices, and they have nonzero amplitudes during the evolution. This contrast can also be seen in the initial states \eqref{eq:s_largeN} and \eqref{eq:s_largeM}.

Continuing, the success probability at time $t$ is the sum of the norm-squares of the amplitudes of $\ket{aa}$, $\ket{ab}$, and $\ket{ac}$, which is
\begin{align}
	p(t) 
		&= \frac{\ell}{(\ell+1)^2} [\cos(\alpha t) - 1]^2 + \frac{k}{2(\ell+1)} \sin^2(\alpha t) \nonumber \\
		&\quad + \frac{1-k}{2(\ell+1)} \sin^2(\alpha t) \nonumber \\
		&= \frac{8\ell\sin ^4(\frac{\alpha t}{2})+(\ell+1) \sin^2(\alpha t)}{2 (\ell+1)^2}. \label{eq:prob_largeM}
\end{align}
This is the same success probability as \eqref{eq:prob}, and so asymptotically, it evolves just like the homogenous lackadaisical quantum walk where each vertex has a self-loop of weight $\ell$. Note although the success probability for small $M$ in \eqref{eq:prob_largeN} is the same as the success probability for large $M$ in \eqref{eq:prob_largeM}, the amplitudes that contribute to each success probability are different. In \eqref{eq:prob_largeN}, success comes from the $\ket{aa}$ and $\ket{ac}$ terms, while in \eqref{eq:prob_largeM}, success comes from the $\ket{aa}$, $\ket{ab}$, and $\ket{ac}$ terms. This is another example of the contribution, or lack thereof, from $b$ vertices.


\subsection{\label{sec:anotherinitial}Another Initial State}

The initial state \eqref{eq:s} that we have used so far is a uniform superposition over the vertices, meaning if we were to measure the position of the walker at the start, we would get each vertex with equal probability. This reflects our initial lack of knowledge of where the marked vertex is, and that each vertex is equally likely to be marked. If we perform a nonhomogeneous lackadaisical quantum walk by applying $SC$ without the query $Q$, however, then the state evolves, even though we have not learned any information about where the marked vertex may be because we have not queried the oracle. To address this, the 1-eigenvector of $U_\text{walk} = SC$ can be used as the starting state instead \cite{Wong31}:
\begin{align}
	\ket{\sigma} 
	    &= \frac{1}{\sqrt{N(N-1) + M\ell + (N-M)\ell'}} \nonumber \\
	    &\quad \times \Bigg[ \sum_{i = 1}^{M} \ket{i} \otimes \left( \sum_{j \ne i} \ket{j} + \sqrt{\ell} \ket{i} \right) \nonumber \\
		&\quad\quad\quad+ \sum_{i = M+1}^{N} \ket{i} \otimes \left( \sum_{j \ne i} \ket{j} + \sqrt{\ell'} \ket{i} \right) \Bigg]. \label{eq:sigma}
\end{align}
While this is not a uniform superposition over the vertices, it has the property that it is unchanged when we walk without the oracle query, i.e., $U_\text{walk} \ket{\sigma} = SC \ket{\sigma} = \ket{\sigma}$. For the irregular complete bipartite graph, such an initial state can lead to a different evolution \cite{Wong31}. For the complete graph, however, we will now prove that the evolution is asymptotically the same, so it does not matter if we use \eqref{eq:s} or \eqref{eq:sigma} as the initial state.

To begin the proof, in \eqref{eq:sigma}, the denominator of the overall factor, for large $N$, is
\[ N(N-1) + M\ell + (N-M)\ell' \approx N^2. \]
Then, for large $N$, \eqref{eq:sigma} asymptotically approaches
\begin{align}
	& \frac{1}{N} \Bigg[ \sum_{i = 1}^{M} \ket{i} \otimes \left( \sum_{j \ne i} \ket{j} + \sqrt{\ell} \ket{i} \right) \label{eq:initial_largeN} \\
		&\quad+ \sum_{i = M+1}^{N} \ket{i} \otimes \left( \sum_{j \ne i} \ket{j} + \sqrt{\ell'} \ket{i} \right) \Bigg]. \nonumber
\end{align}
Next, consider \eqref{eq:s}. Its radicands are, for large $N$,
\begin{gather*}
	N + \ell - 1 \approx N, \\
	N + \ell' - 1 \approx N.
\end{gather*}
Then, for large $N$, \eqref{eq:s} also asymptotically appreaches \eqref{eq:initial_largeN}. Since the two initial states \eqref{eq:s} and \eqref{eq:sigma} both approach \eqref{eq:initial_largeN}, they are asymptotically equivalent. Our numerical simulations are consistent with this; using either initial state results in roughly the same evolution.


\section{Additional Graphs}

In this section, we explore search on a variety of vertex-transitive graphs. Vertex-transitive graphs are necessarily regular, meaning each vertex has the same degree, or number of neighbors. Ignoring self-loops, we denote the degree $d$. For example, the complete graph has a degree $d = N - 1$, since each vertex is adjacent to each of the $N-1$ other vertices. Using a homogeneous lackadaisical quantum walk, the optimal value of $\ell$ for vertex-transitive graphs is asymptotically $d/N$ \cite{Wong28}. For example, for the complete graph, this was $(N-1)/N \approx 1$ for large $N$. Many of the results in this section are very similar to the complete graph from the previous section. Periodic lattices, however, are different, and they suggest that the random weights should be small compared to the degree of the graph.

\begin{figure}
\begin{center}
	\includegraphics{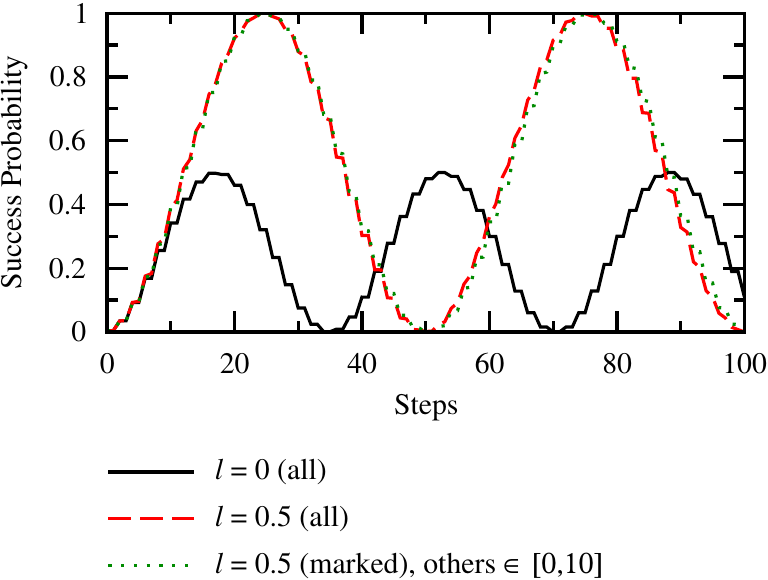}
	\caption{\label{fig:bipartite_128_128}Search on the regular complete bipartite graph with $N = 256$ vertices, which has degree $128$.}
\end{center}
\end{figure}

We begin with the regular complete bipartite graph of $N$ vertices, which consists of two partite sets, each with $N/2$ vertices, such that each vertex is adjacent to every vertex in the other partite set and nonadjacent to every vertex in its own set. So, the degree is $d = N/2$. Search with $N = 256$ is shown in \fref{fig:bipartite_128_128}. The solid black curve is without self-loops \cite{Wong31}. With each self-loop weight equal to the optimal value of $d/N = 1/2$, we get the dashed red curve. Randomly choosing the self-loops at the unmarked vertices to have weights between 0 and 10, which breaks the symmetries of the graph, we get the dotted green curve, and it closely matches the dashed red curve, indicating that only the weight at the marked vertex matters, asymptotically.

\begin{figure}
\begin{center}
	\includegraphics{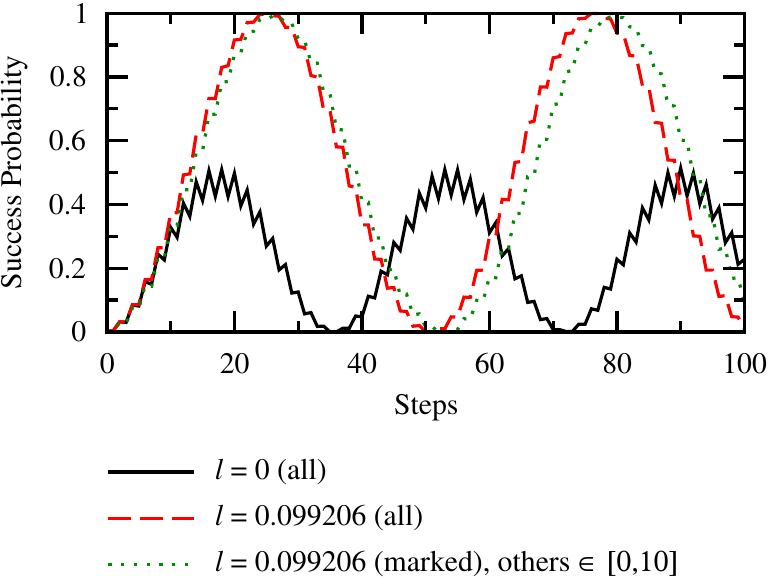}
	\caption{\label{fig:johnson_10_5}Search on the Johnson graph $J(10,5)$, which has $N = 252$ vertices and degree $25$.}
\end{center}
\end{figure}

Johnson graphs are next. A Johnson graph is denoted $J(n,k)$. Its vertices are $k$-element subsets of $n$ symbols, and vertices are adjacent if they differ in exactly one symbol. For example, $J(4,2)$ has four symbols. Using $a$, $b$, $c$, and $d$ as the symbols, the vertices are $ab$, $ac$, $ad$, $bc$, $bd$, and $cd$. Vertices $ab$ and $ac$ are adjacent because they differ in one symbol, and $ab$ and $cd$ are nonadjacent because they differ in two symbols. In general, the Johnson graph $J(n,k)$ has ``$n$~choose~$k$''~$= C(n,k)$ vertices, and each vertex has $k(n-k)$ neighbors. The solid black curve in \fref{fig:johnson_10_5} shows the success probability for search on $J(10,5)$ without self-loops. With each self-loop weight equal to the optimal value of $d/N = k(n-k) / C(n,k) = 25/252 = 0.099206$, we get the dashed red curve. Breaking the symmetries of the graph, in the dotted green curve, we keep the weight of the self-loop at the marked vertex equal to $0.099206$, while the remaining self-loops have weights chosen uniformly at random in the interval $[0,10]$. We see that the success probability evolves similarly to the dashed red curve, so only the weight at the marked vertex matters, asymptotically.

\begin{figure}
\begin{center}
	\includegraphics{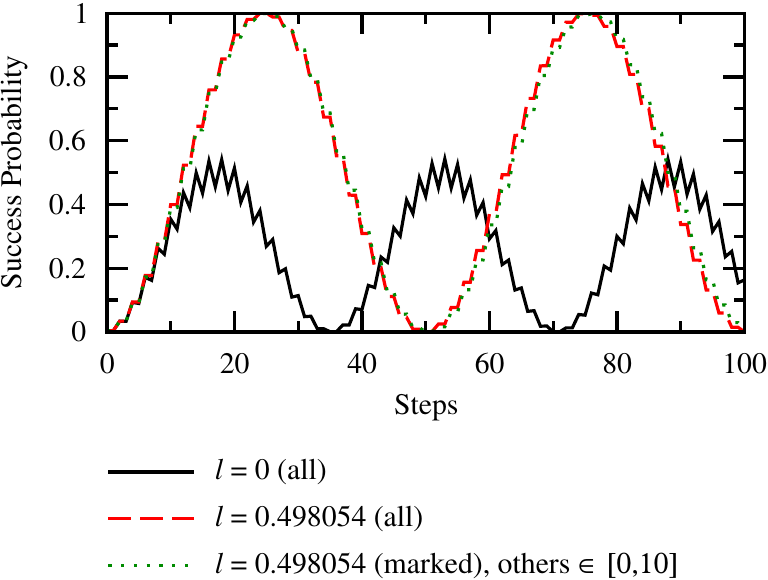}
	\caption{\label{fig:paley_257}Search on the Paley graph of $N = 257$ vertices, which has degree $128$.}
\end{center}
\end{figure}

Strongly regular graphs are also vertex transitive. A strongly regular graph has parameters $(N,d,\lambda,\mu)$, where the graph has $N$ vertices, every vertex has $d$ neighbors, adjacent vertices share $\lambda$ common neighbors, and nonadjacent vertices share $\mu$ common neighbors. One family of strongly regular graphs is the Paley graphs, where $N$ is a prime power such that $N = 1 \pmod 4$. Then, $k = (N-1)/2$, $\lambda = (N-5)/4$, and $\mu = (N-1)/4$. For example, search on the Paley graph (257, 128, 63, 64) is shown in \fref{fig:paley_257}. The solid black curve is the loopless case. With each self-loop weight equal to the optimal value of $d/N = 128/257 = 0.498054$, we get the dashed red curve. Choosing the weights of the self-loops at the unmarked vertices uniformly at random in the interval $[0,10]$, we get the dotted green curve, and it closely matches the dashed red curve. Again, only the weight at the marked vertex matters, asymptotically.

\begin{figure}
\begin{center}
	\subfloat[] {
		\label{fig:lattice_2_16}
		\includegraphics{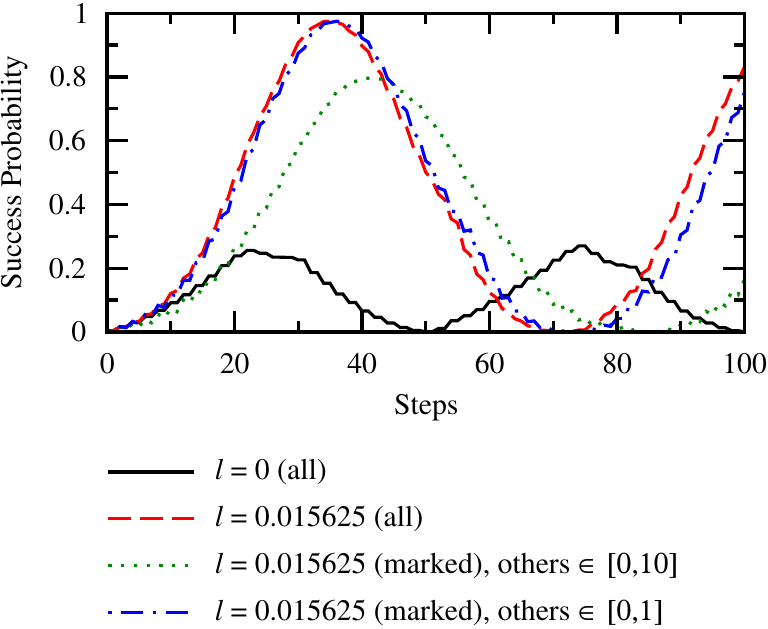}
	}

	\subfloat[] {
		\label{fig:lattice_2_32}
		\includegraphics{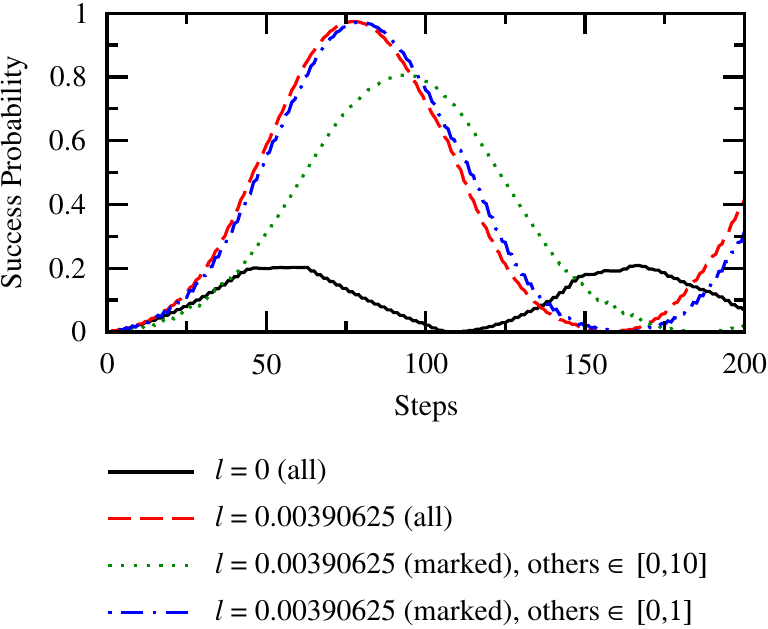}
	}

	\subfloat[] {
		\label{fig:lattice_5_4}
		\includegraphics{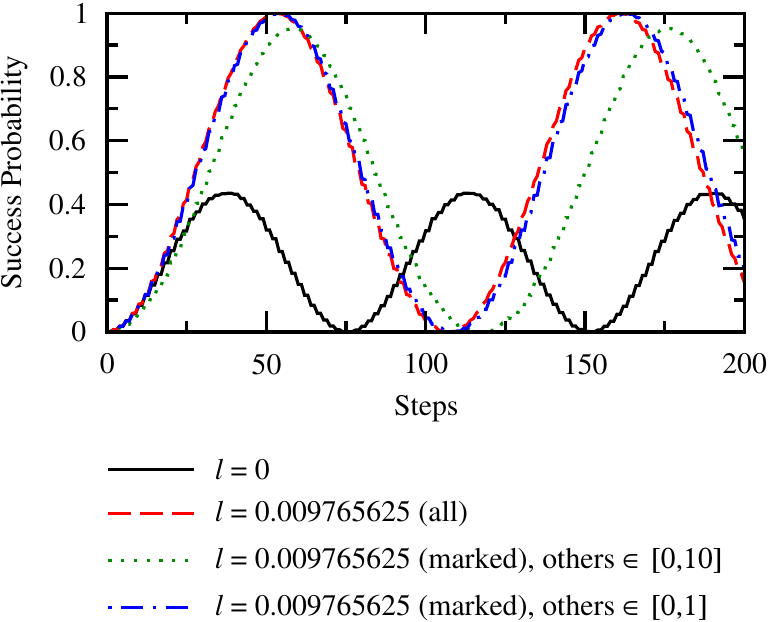}
	}
	\caption{Search on periodic square lattices of various sizes and dimensions. (a) $N = 16 \times 16 = 256$, degree 4. (b) $N = 32 \times 32 = 1024$, degree 4. (c) $N = 4 \times 4 \times 4 \times 4 \times 4 = 1024$, degree 10.}
\end{center}
\end{figure}

Next, we explore search on arbitrary-dimensional periodic square lattices, which will lead to a new observation. \fref{fig:lattice_2_16} shows the success probability for searching the 2D periodic square lattice with $N = 16 \times 16 = 256$ vertices and degree $d = 4$. The solid black curve is the evolution without self-loops \cite{AKR2005}. With each self-loop weight equal to the optimal value of $d/N = 4/N = 4/256 = 0.015625$, we get the dashed red curve. Now, we break the symmetries by only giving the self-loop at the marked vertex a weight of $0.015625$, while the remaining 255 self-loops have weights that are chosen uniformly at random in the interval $[0,10]$. The success probability is shown in the dotted green curve. It does not match the dashed red curve. One might assume this is because $N$ is too small, and so we search a larger 2D lattice in \fref{fig:lattice_2_32} with $N = 32 \times 32 = 1024$ vertices and degree $d = 4$. Again, the dotted green curve differs from the dashed red curve. Thus, increasing $N$ did not help, and search on even larger 2D lattices with $N = 4096$, $16384$, and $65536$ vertices, all of which have degree $4$, confirms that increasing $N$ is not the solution. Instead, let us choose the random self-loops to be in the interval $[0,1]$ so that they are smaller. Now, from the dot-dashed blue curves of \fref{fig:lattice_2_16} and \fref{fig:lattice_2_32}, we get good agreement with the dashed red curves. This suggests that the weights of the random self-loops must be small relative to some quantity, and that quantity is not $N$.

We propose that the quantity is the degree $d$ of the graph. That is, the self-loops at the unmarked vertices do not matter as long as their weights are small compared to the degree of the graph. As intuition, consider an unmarked vertex. If its self-loop has a small weight compared to the number of other edges it has, then the self-loop plays a negligible role in the evolution. The evolution at the vertex is overwhelmingly dictated by the numerous other edges. Next, we offer several tests of this argument.

As a test of our observation, in \fref{fig:lattice_5_4}, we explore search on the 5D lattice of $N = 1024$ vertices, which has degree $d = 10$. Since this has a larger degree than the 2D lattices in \fref{fig:lattice_2_16} and \fref{fig:lattice_2_32}, the self-loops at the unmarked vertices should be less relevant. \fref{fig:lattice_5_4} indicates that this is true. With weights randomly chosen in $[0,10]$, the dotted green curve is closer to the dashed red curve than in \fref{fig:lattice_2_16} and \fref{fig:lattice_2_32}. It is not perfect, however, as some of the weights in $[0,10]$ may be comparable to the degree $d = 10$. If we instead choose the weights in $[0,1]$, then the agreement should be even better, as we confirm in the dotted blue curve of \fref{fig:lattice_5_4}.

In a way, lattices are an outlier because the number of vertices $N$ can be increased without changing the degree $d$. With the previous graphs (the complete graph, regular complete bipartite graphs, Johnson graphs, and Paley graphs), increasing $N$ also increased the degree $d$. So, taking $N$ to be large also takes $d$ to be large, so increasing $N$ will cause the self-loops at unmarked vertices to be irrelevant. Our observation is also consistent with the homogeneous lackadaisical quantum walk, for which $\ell = d/N$ is optimal. With this choice, the self-loops at the unmarked vertices are guaranteed to be small compared to the degree because we are dividing the degree by the number of vertices.

\begin{figure}
\begin{center}
	\subfloat[] {
		\label{fig:hypercube_8}
		\includegraphics{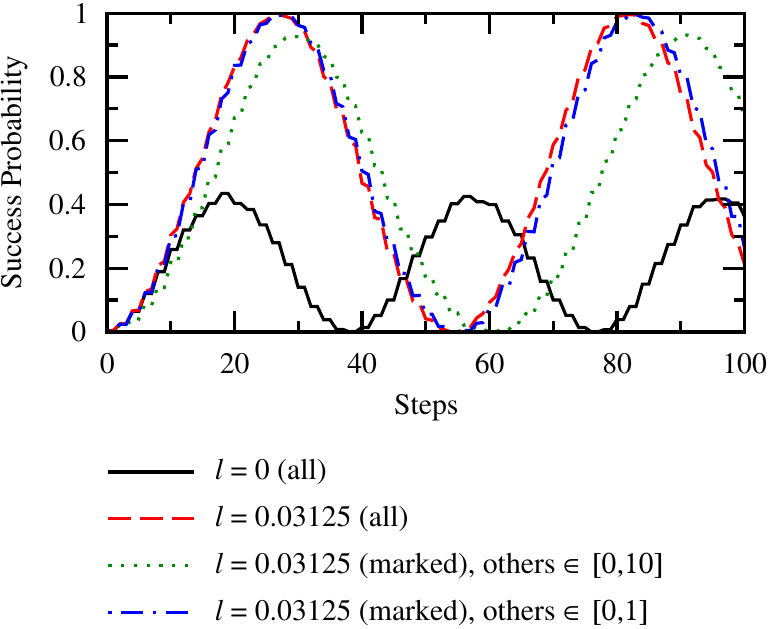}
	}

	\subfloat[] {
		\label{fig:hypercube_14}
		\includegraphics{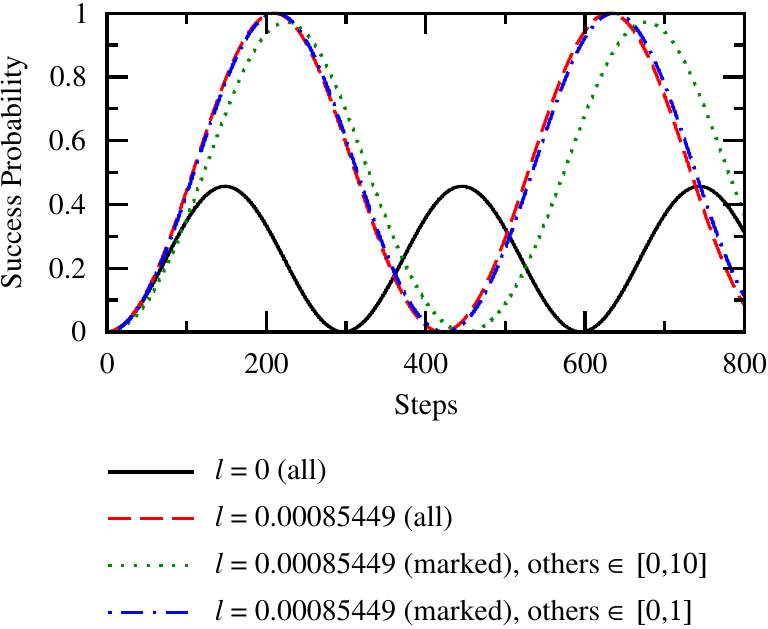}
	}
	\caption{Search on the (a) 10D hypercube, which has $N = 1024$ vertices and degree $10$, and (b) 14D hypercube, which has $N = 16384$ vertices and degree $14$.}
\end{center}
\end{figure}

Next, we consider hypercubes. In 1D, the hypercube is a path of 2 vertices. In 2D, it is a square of 4 vertices. In 3D, it is a cube of 8 vertices. In 4D, it is a tesseract of 16 vertices. In general, the $n$D hypercube has $N = 2^n$ vertices and degree $n$. Search on the 8D hypercube is shown in \fref{fig:hypercube_8}, and the solid black curve is the loopless case. With each self-loop weight equal to the optimal value of $n/2^n = 8/256 = 0.03125$, we get the dashed red curve. Breaking the symmetries of the graph, we keep the weight of the self-loop at the marked vertex equal to $0.03125$, but choose the rest uniformly at random in the interval $[0,10]$. This is the dotted green curve. As we saw with the lattices, it does not match the dashed red curve because the degree $d = 8$ does not dominate the loops chosen in $[0,10]$. If we instead randomly choose the weights in the interval $[0,1]$, we get the dot-dashed blue curve, and this does match the dashed red curve. We could also increase the size of the hypercube, which increases its degree. In \fref{fig:hypercube_14}, we show search on the 14D hypercube, which has $N = 16384$ vertices and degree $d = 14$. The dotted green curve is closer to the dashed red curve, and this is consistent with our observation; by increasing the degree from 8 to 14, the weights at the unmarked vertices become less relevant. As before, choosing the weights in $[0,1]$, as shown in the dot-dashed blue curve, results in even better agreement with the dashed red curve.


\section{Conclusion}

We have shown that a lackadaisical quantum walk can break the symmetries of vertex-transitive graphs while maintaining its speedup for spatial search. That is, its speedup is not dependent on supporting the symmetries of the graph. We demonstrated this by giving every self-loop a different weight, which causes each vertex to evolve differently. Only the weight of the self-loop at the marked vertex affects the search algorithm, as long as the other weights are small compared to the degree of the graph. We proved this for the complete graph for the specific case of two weights, and proving this in general for the complete graph and all vertex-transitive graphs is an open question.

Naturally, one may ask whether a similar result holds for graphs that are not vertex transitive. Prior work on the irregular complete bipartite graph partially answers this. In \cite{Wong32}, the self-loops in one partite set had weight $\ell_1$, while the self-loops in the other partite set had weight $\ell_2$. If the marked vertices are all in the first partite set (with self-loops $\ell_1$), then $\ell_2$ may or may not affect the evolution, depending on the initial state. With the usual initial state that is a uniform superposition over the vertices, then $\ell_2$ can affect the evolution, as shown in Fig.~4 of \cite{Wong32}. If the initial state is a 1-eigenvector of the quantum walk (similar to $\ket{\sigma}$ in Sec.~\ref{sec:anotherinitial}), then $\ell_2$ does not make a difference. Both of these results assume that the number of vertices in each partite set is large. If this is not true, and $\ell_2$ is comparable to or large compared to the number of vertices, then it can affect the evolution with either initial state, as shown in Fig.~3(c) and Fig.~5(c) of \cite{Wong32}.


\begin{acknowledgments}
	This work was supported by startup funds from Creighton University.
\end{acknowledgments}


\appendix

\section{\label{appendix:largeN}Complete Graph, Large $N$}

To find the eigenvectors and eigenvalues of $U$ \eqref{eq:U} for large $N$, we use degenerate perturbation theory. First, we take the leading-order terms of $U$, which gives us the matrix $U_0$:
\[ U_0 = \begin{pmatrix}
	1 & 0 & 0 & 0 & 0 & 0 & 0 & 0 & 0 \\
	0 & 0 & 0 & -1 & 0 & 0 & 0 & 0 & 0 \\
	0 & 0 & 0 & 0 & 0 & 0 & -1 & 0 & 0 \\
	0 & 1 & 0 & 0 & 0 & 0 & 0 & 0 & 0 \\
	0 & 0 & 0 & 0 & -1 & 0 & 0 & 0 & 0 \\
	0 & 0 & 0 & 0 & 0 & 0 & 0 & -1 & 0 \\
	0 & 0 & -1 & 0 & 0 & 0 & 0 & 0 & 0 \\
	0 & 0 & 0 & 0 & 0 & 1 & 0 & 0 & 0 \\
	0 & 0 & 0 & 0 & 0 & 0 & 0 & 0 & 1 
\end{pmatrix}. \]
The normalized eigenvectors and eigenvalues of $U_0$ are much easier to find. They are
\begin{align*}
	& \ket{v_1} = \frac{1}{\sqrt{2}}[0,0,1,0,0,0,1,0,0]^\intercal, \quad -1, \\
	& \ket{v_2} = [0,0,0,0,1,0,0,0,0]^\intercal, \quad -1, \\
	& \ket{v_3} = \frac{1}{\sqrt{2}}[0,0,0,0,0,i,0,1,0]^\intercal, \quad i, \\
	& \ket{v_4} = \frac{1}{\sqrt{2}}[0,i,0,1,0,0,0,0,0]^\intercal, \quad i, \\
	& \ket{v_5} = \frac{1}{\sqrt{2}}[0,0,0,0,0,-i,0,1,0]^\intercal, \quad -i, \\
	& \ket{v_6} = \frac{1}{\sqrt{2}}[0,-i,0,1,0,0,0,0,0]^\intercal, \quad -i, \\
	& \ket{v_7} = [0,0,0,0,0,0,0,0,1]^\intercal, \quad 1, \\
	& \ket{v_8} = \frac{1}{\sqrt{2}}[0,0,-1,0,0,0,1,0,0]^\intercal, \quad 1, \\
	& \ket{v_9} = [1,0,0,0,0,0,0,0,0]^\intercal, \quad 1.
\end{align*}

Next, we lift the degeneracy by including the next-leading-order terms of $U$, which acts as a perturbation. Together, the leading- and next-leading-order terms of $U$ are
\begin{widetext}
\[ U' = \begin{pmatrix}
	1 & 0 & -\frac{2 \sqrt{\ell}}{\sqrt{N}} & 0 & 0 & 0 & 0 & 0 & 0 \\
	0 & 0 & 0 & -1 & 0 & \frac{2}{\sqrt{N}} & 0 & 0 & 0 \\
	0 & 0 & 0 & 0 & 0 & 0 & -1 & 0 & \frac{2}{\sqrt{N}} \\
	0 & 1 & -\frac{2 \sqrt{M_1}}{\sqrt{N}} & 0 & 0 & 0 & 0 & 0 & 0 \\
	0 & 0 & 0 & 0 & -1 & \frac{2 \sqrt{M_\ell}}{\sqrt{N}} & 0 & 0 & 0 \\
	0 & 0 & 0 & 0 & 0 & 0 & 0 & -1 & \frac{2 \sqrt{M_1}}{\sqrt{N}} \\
	-\frac{2 \sqrt{\ell}}{\sqrt{N}} & -\frac{2 \sqrt{M_1}}{\sqrt{N}} & -1 & 0 & 0 & 0 & 0 & 0 & 0 \\
	0 & 0 & 0 & \frac{2}{\sqrt{N}} & \frac{2 \sqrt{M_\ell}}{\sqrt{N}} & 1 & 0 & 0 & 0 \\
	0 & 0 & 0 & 0 & 0 & 0 & \frac{2}{\sqrt{N}} & \frac{2 \sqrt{M_1}}{\sqrt{N}} & 1 
\end{pmatrix}. \]
\end{widetext}
For large $N$, the asymptotic eigenvectors of $U'$ are linear combinations of the degenerate eigenvectors of $U_0$ \cite{Griffiths2018}. For example, starting with the two eigenvectors of $U_0$ that have eigenvalue $-1$, two linear combinations $\alpha_1 \ket{v_1} + \alpha_2 \ket{v_2}$ will be asymptotic eigenvectors of $U'$, i.e.,
\[
       	\begin{pmatrix}
	       	U'_{11} & U'_{12} \\
		U'_{21} & U'_{22} \\
	\end{pmatrix} \begin{pmatrix}
		\alpha_1 \\
		\alpha_2 \\
	\end{pmatrix} = \lambda \begin{pmatrix}
		\alpha_1 \\
		\alpha_2 
	\end{pmatrix}, 
\]
where $U'_{ij} = \bra{v_i} U' \ket{v_j}$, and $\lambda$ is the eigenvalue. Evaluating the matrix components,
\[
	\begin{pmatrix}
		-1 & 0 \\
		0 & -1 \\
	\end{pmatrix} \begin{pmatrix}
		\alpha_1 \\
		\alpha_2 \\
	\end{pmatrix} = \lambda \begin{pmatrix}
		\alpha_1 \\
		\alpha_2 
	\end{pmatrix}.
\]
Solving this eigenvalue relation, two asymptotic eigenvectors and eigenvalues of $U'$ are
\begin{align*}
	& \ket{\Psi_1}=\ket{v_2}, \quad \lambda_1= -1, \\
	& \ket{\Psi_2}=\ket{v_1}, \quad \lambda_2=  -1.
\end{align*}
Similarly, for the eigenvectors of $U_0$ with eigenvalue $i$, two asymptotic eigenvectors of $U'$ take the form $\alpha_3\ket{v_3}+\alpha_4\ket{v_4}$, where
\[    
	\begin{pmatrix}
		U'_{33} & U'_{34} \\
		U'_{43} & U'_{44}\\
	\end{pmatrix}\begin{pmatrix}
		\alpha_3\\
		\alpha_4\\
	\end{pmatrix}=\lambda\begin{pmatrix}
		\alpha_3\\
		\alpha_4
	\end{pmatrix}.
\]
This can be evaluated to get:
\[    
	\begin{pmatrix}
		i & \frac{1}{\sqrt{N}} \\
		\frac{1}{\sqrt{N}} & i\\
	\end{pmatrix}\begin{pmatrix}
		\alpha_3\\
		\alpha_4\\
	\end{pmatrix}=\lambda\begin{pmatrix}
		\alpha_3\\
		\alpha_4
	\end{pmatrix}.
\]
Solving this, two (unnormalized) asymptotic eigenvectors and eigenvalues of $U'$ are: 
\begin{align*}
	& \ket{\Psi_3}=-\ket{v_3}+\ket{v_4}, \quad \lambda_3 = i - \frac{1}{\sqrt{N}}, \\
	& \ket{\Psi_4}=\ket{v_3}+\ket{v_4}, \quad \lambda_4 = i + \frac{1}{\sqrt{N}}. 
\end{align*}
Next, for the eigenvectors of $U_0$ with eigenvalue $-i$, two asymptotic eigenvectors of $U'$ take the form $\alpha_5\ket{v_5}+\alpha_6\ket{v_6}$, where
\[    
	\begin{pmatrix}
		U'_{55} & U'_{56} \\
		U'_{65} & U'_{66}\\
	\end{pmatrix}\begin{pmatrix}
		\alpha_5\\
		\alpha_6\\
	\end{pmatrix}=\lambda\begin{pmatrix}
		\alpha_5\\
		\alpha_6
	\end{pmatrix}.
\]
This can be evaluated to get:
\[
	\begin{pmatrix}
		-i & \frac{1}{\sqrt{N}} \\
		\frac{1}{\sqrt{N}} & -i\\
	\end{pmatrix}\begin{pmatrix}
		\alpha_5\\
		\alpha_6\\
	\end{pmatrix}=\lambda\begin{pmatrix}
		\alpha_5\\
		\alpha_6
	\end{pmatrix}.
\]
Solving this yields the following (unnormalized) asymptotic eigenvectors and eigenvalues of $U'$:
\begin{align*}
	& \ket{\Psi_5}=-\ket{v_5}+\ket{v_6}, \quad \lambda_5 = -i - \frac{1}{\sqrt{N}} \\
	& \ket{\Psi_6}=\ket{v_5}+\ket{v_6}, \quad \lambda_6 = -i + \frac{1}{\sqrt{N}}. 
\end{align*}
Lastly, for eigenvectors of $U_0$ with eigenvalue $1$, three asymptotic eigenvectors of $U'$ take the form $\alpha_7\ket{v_7}+\alpha_8\ket{v_8}+\alpha_9\ket{v_9}$, where
\[    
	\begin{pmatrix}
		U'_{77} & U'_{78} & U'_{79}\\
		U'_{87} & U'_{88} & U'_{89}\\
	      	U'_{97} & U'_{98} & U'_{99}\\
	\end{pmatrix}\begin{pmatrix}
	    	\alpha_7\\
	     	\alpha_8\\
	     	\alpha_9\\
       	\end{pmatrix}=\lambda\begin{pmatrix}
	   	\alpha_7\\
	   	\alpha_8\\
	    	\alpha_9
      	\end{pmatrix}.
\]
This can be evaluated to get:
\[
	\begin{pmatrix}
	    	1 & \sqrt{\frac{2}{N}} & 0\\
	    	-\sqrt{\frac{2}{N}} & 1 & -\sqrt{\frac{2\ell}{N}}\\
	    	0 & \sqrt{\frac{2\ell}{N}} & 1\\
	\end{pmatrix}\begin{pmatrix}
	    	\alpha_7\\
	   	\alpha_8\\
	   	\alpha_9\\
       	\end{pmatrix}=\lambda\begin{pmatrix}
	   	\alpha_7\\
	   	\alpha_8\\
	   	\alpha_9
       	\end{pmatrix}.
\]
Solving this yields the following (unnormalized) asymptotic eigenvectors and eigenvalues of $U'$:
\begin{align*}
	\ket{\Psi_7} 
		&= \frac{1}{\sqrt{\ell}}\ket{v_7}-i\sqrt{\frac{\ell+1}{\ell}}\ket{v_8}+\ket{v_9},\\
		&\quad \lambda_7= 1-i\sqrt\frac{2(\ell+1)}{N} \approx e^{-i\alpha}, \\
	\ket{\Psi_8} 
		&= \frac{1}{\sqrt{\ell}}\ket{v_7}+i\sqrt{\frac{\ell+1}{\ell}}\ket{v_8}+\ket{v_9}, \\
		&\quad \lambda_8= 1+i\sqrt\frac{2(\ell+1)}{N} \approx e^{i\alpha}, \\
	\ket{\Psi_9} 
		&= -\sqrt{\ell}\ket{v_7}+\ket{v_9}, \quad \lambda_9 = 1.
\end{align*}
where $\alpha$ is defined in \eqref{eq:alpha}. Then, by plugging in the respective $\ket{v_i}$, we get the (unnormalized) asymptotic eigenvectors of $U'$ in the $\{ \ket{aa},\ket{ab}, \dots, \ket{cc} \}$ basis that were given in \eqref{eq:eigenvectors_largeN}.


\section{\label{appendix:largeM}Complete Graph, Large $N$ and $M$}

Assuming $M = kN$ for some constant $k$, then for large $N$, the leading-order terms of the search operator $U$ \eqref{eq:U} are
\begin{widetext}
\[ U_0 = \begin{pmatrix}
        1 & 0 & 0 & 0 & 0 & 0 & 0 & 0 & 0 \\
        0 & 0 & 0 & -1 & 0 & 0 & 0 & 0 & 0 \\
        0 & 0 & 0 & 0 & 0 & 0 & -1 & 0 & 0 \\
	0 & 1-2k & -2 \sqrt{k(1-k)} & 0 & 0 & 0 & 0 & 0 & 0 \\
	0 & 0 & 0 & 0 & 2k-1 & 2 \sqrt{k(1-k)} & 0 & 0 & 0 \\
	0 & 0 & 0 & 0 & 0 & 0 & 0 & 2k-1 & 2 \sqrt{k(1-k)} \\
	0 & -2 \sqrt{k(1-k)} & 2k-1 & 0 & 0 & 0 & 0 & 0 & 0 \\
	0 & 0 & 0 & 0 & 2 \sqrt{k(1-k)} & 1-2k & 0 & 0 & 0 \\
	0 & 0 & 0 & 0 & 0 & 0 & 0 & 2 \sqrt{k(1-k)} & 1-2k \\
\end{pmatrix}. \]
The normalized eigenvectors and eigenvalues of $U_0$ are
\begin{align*}
	& \ket{v_1} = \left[0,0,0,0,1-k,-\sqrt{k(1-k)},0,-\sqrt{k(1-k)},k\right]^\intercal, \quad -1, \\
	& \ket{v_2} = \frac{1}{\sqrt{2}} \left[0,\sqrt{k},\sqrt{1-k},\sqrt{k},0,0,\sqrt{1-k},0,0\right]^\intercal, \quad -1, \\
	& \ket{v_3} = \left[0,0,0,0,-\sqrt{k(1-k)},k - \frac{1}{2} - \frac{i}{2},0,k-\frac{1}{2}+\frac{i}{2},\sqrt{k(1-k)}\right]^\intercal, \quad i, \\
	& \ket{v_4} = \frac{1}{\sqrt{2}} \left[0,-i\sqrt{1-k},i \sqrt{k},-\sqrt{1-k},0,0,\sqrt{k},0,0\right]^\intercal, \quad i, \\
	& \ket{v_5} = \left[0,0,0,0,-\sqrt{k(1-k)},k-\frac{1}{2}+\frac{i}{2},0,k-\frac{1}{2}-\frac{i}{2},\sqrt{k(1-k)}\right]^\intercal, \quad -i, \\
	& \ket{v_6} = \frac{1}{\sqrt{2}} \left[0,i\sqrt{1-k},-i \sqrt{k},-\sqrt{1-k},0,0,\sqrt{k},0,0\right]^\intercal, \quad -i, \\
	& \ket{v_7} = \left[0,0,0,0,k,\sqrt{k(1-k)},0,\sqrt{k(1-k)},1-k\right]^\intercal, \quad 1, \\
	& \ket{v_8} = \frac{1}{\sqrt{2}} \left[0,-\sqrt{k},-\sqrt{1-k},\sqrt{k},0,0,\sqrt{1-k},0,0\right]^\intercal, \quad 1, \\
	& \ket{v_9} = [1,0,0,0,0,0,0,0,0]^\intercal, \quad 1.
\end{align*}
Adding the next-order terms of the search operator $U$ \eqref{eq:U}, we get
\[ U' = \begin{pmatrix}
	1 & -\frac{2 \sqrt{k\ell}}{\sqrt{N}} & -\frac{2 \sqrt{(1-k)\ell}}{\sqrt{N}} & 0 & 0 & 0 & 0 & 0 & 0 \\
	0 & 0 & 0 & -1 & \frac{2 \sqrt{k}}{\sqrt{N}} & \frac{2 \sqrt{1-k}}{\sqrt{N}} & 0 & 0 & 0 \\
	0 & 0 & 0 & 0 & 0 & 0 & -1 & \frac{2 \sqrt{k}}{\sqrt{N}} & \frac{2 \sqrt{1-k}}{\sqrt{N}} \\
	-\frac{2 \sqrt{k\ell}}{\sqrt{N}} & 1-2k & -2 \sqrt{k(1-k)} & 0 & 0 & 0 & 0 & 0 & 0 \\
	0 & 0 & 0 & \frac{2 \sqrt{k}}{\sqrt{N}} & 2k-1 & 2 \sqrt{k(1-k)} & 0 & 0 & 0 \\
	0 & 0 & 0 & 0 & 0 & 0 & \frac{2 \sqrt{k}}{\sqrt{N}} & 2 k-1 & 2 \sqrt{k(1-k)} \\
	-\frac{2 \sqrt{(1-k)\ell}}{\sqrt{N}} & -2 \sqrt{1-k} \sqrt{k} & 2 k-1 & 0 & 0 & 0 & 0 & 0 & 0 \\
	0 & 0 & 0 & \frac{2 \sqrt{1-k}}{\sqrt{N}} & 2 \sqrt{k(1-k)} & 1-2k & 0 & 0 & 0 \\
	0 & 0 & 0 & 0 & 0 & 0 & \frac{2 \sqrt{1-k}}{\sqrt{N}} & 2 \sqrt{k(1-k)} & 1-2k \\
\end{pmatrix}. \]
\end{widetext}
From degenerate perturbation theory, linear combinations of the degenerate eigenvectors of $U_0$ are asymptotic eigenvectors of $U'$. Using this with the eigenvectors of $U_0$ with eigenvalue $-1$, for large $N$, two linear combinations $\alpha_1 \ket{v_1} + \alpha_2 \ket{v_2}$ will be asymptotic eigenvectors of $U'$, i.e.,
\[
       	\begin{pmatrix}
	       	U'_{11} & U'_{12} \\
		U'_{21} & U'_{22} \\
	\end{pmatrix} \begin{pmatrix}
		\alpha_1 \\
		\alpha_2 \\
	\end{pmatrix} = \lambda \begin{pmatrix}
		\alpha_1 \\
		\alpha_2 
	\end{pmatrix}, 
\]
where $U'_{ij} = \bra{v_i} U' \ket{v_j}$ and $\lambda$ is the eigenvalue. Evaluating the matrix components,
\[
	\begin{pmatrix}
		-1 & 0 \\
		0 & -1 \\
	\end{pmatrix} \begin{pmatrix}
		\alpha_1 \\
		\alpha_2 \\
	\end{pmatrix} = \lambda \begin{pmatrix}
		\alpha_1 \\
		\alpha_2 
	\end{pmatrix}.
\]
Solving this, two asymptotic eigenvectors and eigenvalues of $U'$ are
\begin{align*}
	& \ket{\Psi_1}=\ket{v_2}, \quad \lambda_1= -1, \\
	& \ket{\Psi_2}=\ket{v_1}, \quad \lambda_2=  -1.
\end{align*}
Similarly this can be done for each of the other grouped degenerate eigenvectors of $U_0$. For the eigenvectors of $U_0$ with eigenvalue $i$, two asymptotic eigenvectors of $U'$ take the form $\alpha_3\ket{v_3}+\alpha_4\ket{v_4}$, i.e.,
\[
	\begin{pmatrix}
		U'_{33} & U'_{34} \\
		U'_{43} & U'_{44}\\
	\end{pmatrix}\begin{pmatrix}
		\alpha_3\\
		\alpha_4\\
	\end{pmatrix}=\lambda\begin{pmatrix}
		\alpha_3\\
		\alpha_4
	\end{pmatrix}.
\]
This can be evaluated to get:
\[
	\begin{pmatrix}
		i & \frac{1+i}{\sqrt{2N}} \\
		\frac{1-i}{\sqrt{2N}} & i \\
	\end{pmatrix}\begin{pmatrix}
		\alpha_3\\
		\alpha_4\\
	\end{pmatrix}=\lambda\begin{pmatrix}
		\alpha_3\\
		\alpha_4
	\end{pmatrix}.
\]
Solving this yields the following (unnormalized) asymptotic eigenvectors and eigenvalues of $U'$: 
\begin{align*}
	& \ket{\Psi_3} = \frac{1+i}{\sqrt{2}} \ket{v_3} + \ket{v_4}, \quad \lambda_3 = i + \frac{1}{\sqrt{N}} \approx ie^{-i\phi}, \\
	& \ket{\Psi_4} = -\frac{1+i}{\sqrt{2}} \ket{v_3} + \ket{v_4}, \quad \lambda_4 = i - \frac{1}{\sqrt{N}} \approx ie^{i\phi},
\end{align*}
where
\[ \phi = \sin^{-1} \left( \frac{1}{\sqrt{N}} \right). \]
For the eigenvectors of $U_0$ with eigenvalue $-i$, the linear combination would be $\alpha_5\ket{v_5}+\alpha_6\ket{v_6}$ which can be solved by the matrix:
\[
	\begin{pmatrix}
		U'_{55} & U'_{56} \\
		U'_{65} & U'_{66}\\
	\end{pmatrix}\begin{pmatrix}
		\alpha_5\\
		\alpha_6\\
	\end{pmatrix}=\lambda\begin{pmatrix}
		\alpha_5\\
		\alpha_6
	\end{pmatrix}.
\]
This can be evaluated to get:
\[
	\begin{pmatrix}
		-i & \frac{1-i}{\sqrt{2N}} \\
		\frac{1+i}{\sqrt{2N}} & -i \\
	\end{pmatrix}\begin{pmatrix}
		\alpha_5\\
		\alpha_6\\
	\end{pmatrix}=\lambda\begin{pmatrix}
		\alpha_5\\
		\alpha_6
	\end{pmatrix}.
\]
Solving this matrix then gives us the following (unnormalized) asymptotic eigenvectors and eigenvalues:
\begin{align*}
	& \ket{\Psi_5} = \frac{1-i}{\sqrt{2}} \ket{v_5} + \ket{v_6}, \quad \lambda_5 = -i + \frac{1}{\sqrt{N}} \approx -ie^{i\phi}, \\
	& \ket{\Psi_6} = -\frac{1-i}{\sqrt{2}} \ket{v_5} + \ket{v_6}, \quad \lambda_6 = -i - \frac{1}{\sqrt{N}} \approx -ie^{-i\phi}.
\end{align*}
Lastly, for the eigenvectors of $U_0$ with eigenvalue $1$, the linear combination would be $\alpha_7\ket{v_7}+\alpha_8\ket{v_8}+\alpha_9\ket{v_9}$ which can be solved by the matrix:
\[
	\begin{pmatrix}
		U'_{77} & U'_{78} & U'_{79}\\
		U'_{87} & U'_{88} & U'_{89}\\
	      	U'_{97} & U'_{98} & U'_{99}\\
	\end{pmatrix}\begin{pmatrix}
	    	\alpha_7\\
	     	\alpha_8\\
	     	\alpha_9\\
       	\end{pmatrix}=\lambda\begin{pmatrix}
	   	\alpha_7\\
	   	\alpha_8\\
	    	\alpha_9
      	\end{pmatrix}.
\]
This can be evaluated to get:
\[
	\begin{pmatrix}
		1 & \sqrt{\frac{2}{N}} & 0 \\
		-\sqrt{\frac{2}{N}} & 1 & -\sqrt{\frac{2\ell}{N}} \\
		0 & \sqrt{\frac{2\ell}{N}} & 1 \\
	\end{pmatrix}\begin{pmatrix}
	    	\alpha_7\\
	   	\alpha_8\\
	   	\alpha_9\\
       	\end{pmatrix}=\lambda\begin{pmatrix}
	   	\alpha_7\\
	   	\alpha_8\\
	   	\alpha_9
       	\end{pmatrix}.
\]
Solving this matrix then gives us the following (unnormalized) asymptotic eigenvectors and eigenvalues:
\begin{align*}
	\ket{\Psi_7} 
		&= \frac{1}{\sqrt{\ell}} \ket{v_7} - i \sqrt{\frac{\ell+1}{\ell}} \ket{v_8} + \ket{v_9},\\
		&\quad \lambda_7= 1-\frac{i\sqrt{2(\ell+1)}}{\sqrt{N}} \approx e^{-i\alpha}, \\
	\ket{\Psi_8} 
		&= \frac{1}{\sqrt{\ell}} \ket{v_7} + i \sqrt{\frac{\ell+1}{\ell}} \ket{v_8} + \ket{v_9}, \\
		&\quad \lambda_8= 1+\frac{i\sqrt{2(\ell+1)}}{\sqrt{N}} \approx e^{i\alpha}, \\
	\ket{\Psi_9} 
		&= -\sqrt{\ell}\ket{v_7}+\ket{v_9}, \quad \lambda_9 = 1,
\end{align*}
where $\alpha$ is defined in \eqref{eq:alpha}. Then, plugging in the respective $\ket{v_i}$'s, we get the approximate eigenvectors of $U$ in the $\{ \ket{aa},\ket{ab}, \dots, \ket{cc} \}$ basis, which were given in \eqref{eq:eigenvectors_largeM}.


\bibliography{refs}

\end{document}